\documentclass[12pt]{article}

\usepackage{citesort}
\input{epsf}

\newcommand{\sect}[1]{\setcounter{equation}{0}\section{#1}}

\begin{document}

\title{Lorentzian Condition in Quantum Gravity}
\author{{\sc Raphael Bousso}$^{\rm a}$\thanks{\it
 bousso@stanford.edu} \ 
  and {\sc Stephen W. Hawking}$^{\rm b}$\thanks{\it
 s.w.hawking@damtp.cam.ac.uk}
      \\[1 ex] {\it {\small
\begin{tabular}{ll}
$\!\!\!^{\rm a}$Department of Physics
& $\!\!\!^{\rm b}$DAMTP \\
Stanford University
& University of Cambridge   \\
Stanford, CA 94305-4060 \makebox[3em]{}
& Silver Street, Cambridge CB3 9EW \\
U.S.A.
& United Kingdom \\
\end{tabular} } }
       }
\date{SU-ITP-98-26~~~~DAMTP-1998-87~~~~hep-th/9807148}

\maketitle

\begin{abstract}
  
  The wave function of the universe is usually taken to be a
  functional of the three-metric on a spacelike section, $\Sigma$,
  which is measured.  It is sometimes better, however, to work in the
  conjugate representation, where the wave function depends on a
  quantity related to the second fundamental form of $\Sigma$.  This
  makes it possible to ensure that $\Sigma$ is part of a Lorentzian
  universe by requiring that the argument of the wave function be
  purely imaginary.  We demonstrate the advantages of this formalism
  first in the well-known examples of the nucleation of a de~Sitter or
  a Nariai universe.  We then use it to calculate the pair creation
  rate for sub-maximal black holes in de~Sitter space, which had been
  thought to vanish semi-classically.

\end{abstract}

\pagebreak

\sect{Introduction}

The no boundary proposal~\cite{HarHaw83} is formulated in terms of
Euclidean path integrals. But the world we live in is Lorentzian, or
at least we interpret our observations in terms of Lorentzian
spacetime. One therefore has to continue the results from the
Euclidean path integrals analytically to the Lorentzian regime.

The approach to quantum cosmology that has been followed in the past
is to examine the behavior of the wave function, as a function of the
overall scale, $a$, of the metric, $h_{ij}$, on the spacelike
surface, $\Sigma$.  If the dependence on $a$ was exponential, this was
interpreted as corresponding to a Euclidean spacetime, while an
oscillatory dependence on $a$ was interpreted as corresponding to a
Lorentzian spacetime.

For example, in the case of Einstein gravity with a cosmological
constant $\Lambda$, the path integral for the wave function of a
three-sphere of radius $a$ will be dominated by an instanton which is
part of a four-sphere of radius $R_0 = \sqrt{3/\Lambda}$. In this
saddlepoint approximation, the wave function will be given by
$e^{-I}$, where $I$ is the Euclidean action of the saddlepoint
geometry; we are neglecting a prefactor. For $a<R_0$, there will be a
real Euclidean geometry, bounded by the three-sphere, $\Sigma$, of
radius $a$.  The wave function, $\Psi$, will be 1 for $a =0$, and will
increase rapidly with $a$, up to $a=R_0$. For $a>R_0$, there are no
Euclidean solutions with the given boundary conditions.

There are, however, two complex solutions, each of which can be thought
of as half the Euclidean four-sphere, joined to part of the Lorentzian
de~Sitter solution. The real part of the action of these complex
solutions is equal to the action of the Euclidean half-four-sphere,
and is the same for all values of $a$. On the other hand, the
imaginary part of the action comes from the Lorentzian de~Sitter part
of the solution, and depends on $a$.  Thus the wave function for large
$a$ oscillates rapidly with constant amplitude.

This shows the association between an oscillatory wave function and a
Lorentzian spacetime, but the distinction between exponential and
oscillatory is not precise, and does not identify which part of the
wave function describes which physical situation. In more complicated
situations, the saddlepoint complex solutions will not separate
neatly into Euclidean and Lorentzian parts. So it is not clear how to
calculate the probability of Lorentzian geometries.

One might apply appropriate operators to the wave function to recover
information about whether a given spacelike surface is part of a
Lorentzian or a Euclidean spacetime. But the use of operators is
cumbersome and requires the evaluation of $\Psi$ for a range of
arguments. It would be preferable if the observable geometric
properties, such as the Lorentzian character of the universe, were
manifest in the argument of the wave function. The square of its
amplitude would then yield a probability measure for any given set of
such quantities.

We therefore want to put forward an approach which focuses on the
defining characteristic of a Lorentzian geometry in the neighbourhood
of $\Sigma$.  This is that the induced metric, $h_{ij}$, on $\Sigma$
should be real, but the second fundamental form,
\begin{equation}
K_{ij} = \nabla_i n_j,
\end{equation}
defined for Euclidean signature, should be purely imaginary. Here
$n^j$ is the unit normal to the surface $\Sigma$. The second
fundamental form is also called the extrinsic curvature of the surface
$\Sigma$ in the manifold $M$. It can be regarded as the derivative of
the metric, $h_{ij}$, on $\Sigma$, as $\Sigma$ is moved in its normal
direction in $M$. Thus requiring the second fundamental form to be
purely imaginary means that $h_{ij}$ has a real derivative with
respect to the Lorentzian time coordinate, $t= \mbox{Im}(\tau)$, where
$\tau$ is Euclidean time. This is the condition for a Lorentzian
geometry in a neighbourhood of $\Sigma$.

The second fundamental form, $K_{ij}$, is trivially related to 
$\pi_{ij}$, the momentum conjugate to $h_{ij}$:
\begin{equation}
\pi_{ij} = - h^{1/2} ( K_{ij} - h_{ij} K_{kl} h^{kl} ),
\end{equation}
where $h$ is the determinant of the metric $h_{ij}$.  Clearly, for
real metrics $h_{ij}$, taking $K_{ij}$ to be purely imaginary is
equivalent to taking $\pi_{ij}$ purely imaginary.  It is easy to
transform from the usual representation of the wave function,
$\Psi[h_{ij}]$, to the momentum representation, in which the wave
function is a functional of $\pi_{ij}$.  The two representations are
related by a Laplace transform:
\begin{equation}
\Psi \left[ \pi^{ij} \right] = \int d \left[ h_{ij} \right]
\Psi \left[ h_{ij} \right]
\exp \left( - \int_\Sigma d^3x {\pi^{ij} h_{ij} }\right),
\end{equation}
where the integral over the metric components at each point of
$\Sigma$ is taken to be over all $ h_{ij} $ with positive determinant
$h$. This Laplace transform can be analytically continued to complex
values of $\pi ^{ij} $. The wave function for a universe that is
Lorentzian in a neighbourhood of $\Sigma$ is then obtained by taking
$\pi ^{ij} $ to be purely imaginary.

Thus the requirement that we live in a Lorentzian universe can be made
manifest in the argument of the wavefunction. Further support for
choosing the momentum representation comes from the fact that we
cannot measure the metric globally on a spacelike section, but that
the expansion rate of the universe, which is related to the second
fundamental form, is easily observable.

The saddlepoint approximation to the wave function will be
\begin{equation}
\Psi \left[ \pi^{ij} \right] = e^{-I},
\end{equation}
where we neglect a prefactor;
\begin{equation}
I = -{1\over 16\pi} \int d^4x g^{1/2}(R-2\Lambda )
\label{eq-action}
\end{equation}
is the Euclidean action\footnote{Note that this action does not
  contain the usual surface term, which is cancelled exactly in the
  Laplace transform.} of a complex solution of the field equations
with the imaginary given values of $\pi ^{ij} $ on $\Sigma $. This
complex saddlepoint solution will be Lorentzian near $\Sigma $ by
construction. Further away it may be complex or Euclidean but this
does not matter because one is making measurements only on $\Sigma $.
One therefore has to perform a path integral over the metric
everywhere except on $\Sigma $. The use of a complex saddlepoint
solution does not mean that spacetime is complex. It can just be
regarded as a mathematical trick to evaluate the path integral.

\section{Homogeneous Isotropic Universe without Black Holes}

We can illustrate the above discussion by a consideration of general
relativity without matter fields but with a cosmological constant
$\Lambda $. Because we are not interested in gravitational waves, we
shall restrict ourselves to spherically symmetric solutions. This
means that the second fundamental form $K_{ij}$ has two independent
components, $K_s$ and $K_l$. By a gauge choice, we can consider only
cases with $K_l$ constant on $\Sigma$.

A homogeneous isotropic universe without black holes is the background
with respect to which we have to compare the probability of a universe
containing a pair of black holes. This is the familiar de~Sitter
model, with the Euclidean saddlepoint metric
\begin{equation}
ds^2 = V(r) d\tau^2 + V(r)^{-1} dr^2 + r^2 d\Omega^2,
\end{equation}
where
\begin{equation}
V(r) = 1 - \frac{\Lambda}{3} r^2.
\end{equation}
We can make a choice of coordinates in which the spacelike surfaces
$\Sigma$ will be round three-spheres. Then the metric takes the form
\begin{equation}
ds^2 =  d\hat\tau^2 + a(\hat\tau)^2 d\Omega_3^2,
\end{equation}
where $d\Omega_3^2 = dx^2 + \sin^2 x d\Omega_2^2$ is the metric on the
unit three-sphere, and
\begin{equation}
a(\hat\tau) = R_0 \sin(R_0^{-1} \hat\tau).
\end{equation}

The second fundamental form\footnote{As we pointed out in the previous
  section, we should strictly be working with the canonical momentum,
  $\pi_{ij}$. The Lorentzian condition that the argument of the
  wavefunction be purely imaginary, however, can equally well be
  implemented for various combinations of $\pi_{ij}$ and $h_{ij}$,
  such as $K_{ij}$ or $K_i^{\ j}$. Here we are choosing the latter
  quantity for the sake of clarity, since it leads to rather simple
  equations. It is straightforward to repeat the treatment using
  components of $\pi_{ij}$.}  $K_i^{\ j}$ contains only one
independent component, $K = K_l$, since
\begin{equation}
K_l = K_s = \frac{\dot{a}}{a};
\end{equation}
an overdot denotes differentiation with respect to Euclidean time
$\hat\tau$.  For $K$ real (i.e.\ Euclidean), there will always be a
real Euclidean solution. For positive $K$, this will be less than half
the Euclidean four-sphere of radius $R_0$ and for $K$ negative, it
will be more than half. The action will be
\begin{equation}
I_{\rm dS}(K) = - \frac{3 \pi}{2 \Lambda}
 \left[ 1 - \frac{(3 + 2 K^2) K}{2(1 + K^2)^{3/2}} \right].
\label{eq-act-s3}
\end{equation}
The saddlepoint approximation to the wave function, neglecting
the prefactor $A $, will be
\begin{equation}
\Psi(K) = \exp \left[ - I_{\rm dS}(K) \right] .
\end{equation}
For $K=0$, the saddlepoint solution will be half the Euclidean
four-sphere and the wave function will be
\begin{equation}
\Psi = \exp \left( \frac{3 \pi}{2 \Lambda} \right).
\label{eq-Psi-k=0}
\end{equation}

\begin{figure}[htb]
\epsfxsize=\textwidth
\epsfbox{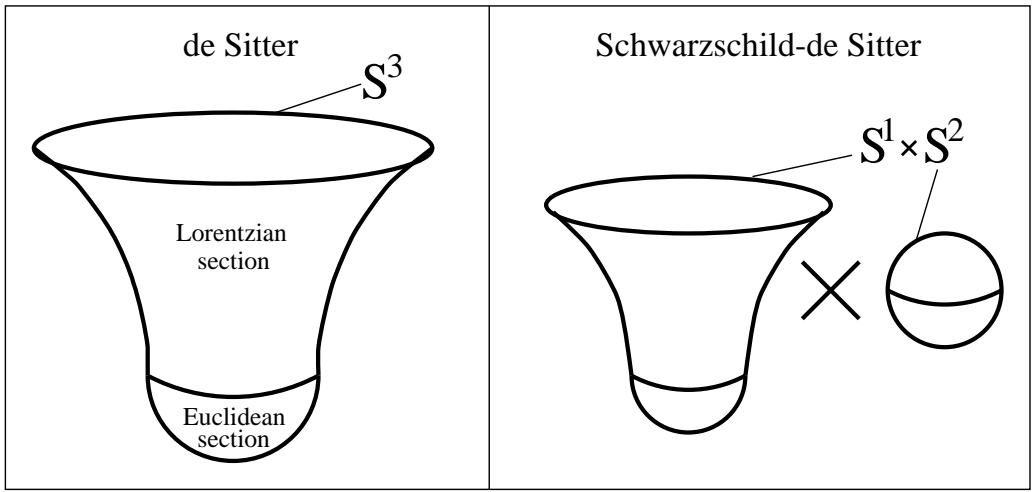}
\caption[Creation of de~Sitter and Schwarzschild-de~Sitter
universes]{\small\sl The creation of a de~Sitter universe (left) can
  be visualized as half of a Euclidean four-sphere joined to a
  Lorentzian four-hyperboloid. The picture on the right shows the
  corresponding nucleation process for a de~Sitter universe containing
  a pair of black holes. In this case the spacelike slices have
  non-trivial topology.}
\label{fig-tun}
\end{figure}

Having calculated the wave function for real $K$, one can now
analytically continue to complex values. Up the imaginary $K$ axis,
only the imaginary part of the action will change, as can be seen from
Eq.~(\ref{eq-act-s3}). Thus, the amplitude of the wave function will
remain at the value for $K=0$ given in Eq.~(\ref{eq-Psi-k=0}). But the
phase of the wave function will vary rapidly with the imaginary part
of $K$. The wave function for positive imaginary $K$ will be be given
by just one of the two complex solutions we had before. It is the one
that consists of the half Euclidean four-sphere, joined to an
expanding de~Sitter solution across a minimal three-sphere spatial
section (see Fig.~\ref{fig-tun}).

Thus this approach separates the expanding and contracting phases of
the de~Sitter universe, which occur when one looks at the wave
function in the $h_{ij}$ representation.  This makes contact with the
tunneling proposal for the wave function~\cite{Vil86} (see
also~\cite{Lin84b} for earlier work). In this one selects the solution
of the Wheeler-DeWitt equation that is {\it outgoing} at large values
of the scale factor $a$. One can regard the Lorentzian condition as a
precise definition of outgoing. However, the probability according to
the tunneling proposal is $e^{+I}$ rather than $e^{-I}$ as with the no
boundary proposal.

\section{Universe with Maximal Black Holes}

To get a universe containing black holes, one would like to calculate
the probability for a Lorentzian geometry on a spacelike surface
$\Sigma$ with $n$ handles. This would represent an expanding universe,
with $n$ pairs of black holes, that inflated from spacetime foam. It
seems reasonable to suppose that the probability of $n$ handles is
roughly the $n$'th power of the probability of a single handle, with
appropriate phase space factors.  Thus it is sufficient to consider
the relative probabilities for zero and one handles. We shall restrict
ourselves to spherical symmetry, to make the problem tractable, but it
is reasonable to assume that spherical configurations dominate the
path integral.

The zero handle surfaces (topology $S^3$) correspond to the Lorentzian
de~Sitter solution, while the one handle surfaces (topology $S^1
\times S^2$) correspond to Schwarzschild-de~Sitter, with the
Lorentzian metric
\begin{equation}
ds^2 = - V(r) dt^2 + V(r)^{-1} dr^2 + r^2 d\Omega^2,
\end{equation}
where
\begin{equation}
V(r) = 1 - \frac{2\mu}{r} - \frac{\Lambda}{3} r^2.
\label{eq-v}
\end{equation}
This represents a pair of black holes in a de~Sitter background. The
mass parameter, $\mu$, of the black holes can be in the range from
zero up to a maximum value of $1/(3 \sqrt{\Lambda})$. For mass less
than the maximum value, the surface gravity of the black hole horizon
is greater than that of the cosmological horizon. This means that if
one tries to turn the Schwarzschild-de~Sitter solution into a compact
Euclidean instanton ($d\tau = i dt$), one gets a conical singularity
either on the black hole horizon or on the cosmological horizon. For
this reason, it has been thought that black holes could spontaneously
nucleate in a de~Sitter background only if they had the maximum
mass~\cite{GinPer83,BouHaw95,BouHaw96}. We shall show in the next
section that this conditions can in fact be relaxed.

For now, we shall focus on the maximal case. In this limit, the
Schwarz\-schild-de~Sitter solution degenerates into the Nariai
solution, in which the two horizons have the same area and surface
gravity, and a compact Euclidean instanton is possible without conical
singularities:
\begin{equation}
ds^2 = d\hat\tau^2 + a(\hat\tau)^2 dx^2 + R_1^2 d\Omega_2^2,
\end{equation}
where $a(\hat\tau) = R_1 \sin(R_1^{-1} \hat\tau)$.  The two-spheres on
$\Sigma$ all have the same radius, $R_1 = 1/\sqrt{\Lambda}$, so
$K_s=0$ and there will be only one independent component of the second
fundamental form, $K = K_l$.  The Euclidean saddlepoint is a direct
product of two round two-spheres of radius $R_1$. The Lorentzian
Nariai solution is the direct product of (1+1)-dimensional de~Sitter
space with a round two-sphere.

The value of $K$ will govern the size of the first Euclidean
two-sphere in the same way it did for the de~Sitter four-sphere in the
previous section. For real $K$, the geometry is entirely Euclidean,
while for imaginary $K$, it will consist of half of $S^2 \times S^2$,
joined to the expanding half of the Lorentzian Nariai solution (see
Fig.~\ref{fig-tun}).
The action will be given by
\begin{equation}
I_{\rm N}(K) = - \frac{\pi}{\Lambda}
 \left( 1 - \frac{K}{\sqrt{1 + K^2}} \right),
\end{equation}
yielding the wave function
\begin{equation}
\Psi_{\rm N}(K) = \exp \left[ I_{\rm N}(K) \right] .
\end{equation}

To obtain a Lorentzian universe, we must choose $K$ to be purely
imaginary.  Then the real part of the Euclidean action, which gives
the amplitude of the wave function, will be $-2\pi/\Lambda$.  As in
the de~Sitter case, this is independent of $K$ as long as ${\rm Re}
(K)=0$. The imaginary part of the action, which gives the phase of the
wave function, depends on $K$.

To calculate the pair creation rate of Nariai black holes on a
de~Sitter background, we note that $\Psi^* \Psi$ is a probability
measure. It is important to stress that the probability measure
depends only on the real part of the saddlepoint action, which stems
from the Euclidean sector. In accordance with other instanton methods,
the pair creation rate $\Gamma_{\rm N}$ can thus be obtained by
normalising this probability with respect to de~Sitter space:
\begin{equation}
\Gamma_{\rm N} =
 \frac{\Psi_{\rm N}^* \Psi_{\rm N}}{\Psi_{\rm dS}^* \Psi_{\rm dS}}
 = \exp \left\{ - 2 \left[ {\rm Re}(I_{\rm N}) - {\rm Re}(I_{\rm dS})
      \right] \right\}
 = \exp \left( \frac{-\pi}{\Lambda} \right).
\end{equation}
Therefore the pair creation of black holes is highly suppressed except
when the (effective) cosmological constant is close to the Planck
value, as it may have been in the earliest stages of inflation.

\section{Universe with Sub-Maximal Black Holes}

In the previous section, we chose to consider only black holes of
maximal size in order to avoid a conical singularity in the Euclidean
saddlepoint solution. For a metric to dominate the path integral, it
has to be a solution of the Einstein equations at every point of the
manifold; but on a conical singularity clearly it is not. Thus the
action will not be stationary with respect to general variations of a
metric containing a conical singularity.

However, conical singularities are expected in general on the
measurement surface $\Sigma$ if one is working in the metric
representation. The solution of the field equations for given $h_{ij}$
on $\Sigma$ will in general have a non-zero second fundamental form
$K_{ij}$ on $\Sigma$. When this solution is joined to its reflection
across $\Sigma$ to calculate $\Psi^* \Psi$, one gets a conical
singularity in general.

The rule is that conical singularities are expected on $\Sigma$ if
they correspond to components of the metric that are measured.  For
example, if one wants the probability of an $S^1 \times S^2$ handle
with a two-sphere cross section, $\sigma$, of area $A$, one can impose
the Lorentzian condition that the real part of the second fundamental
form vanish everywhere on $\Sigma$ except for $\sigma$.  One cannot
specify the second fundamental form on $\sigma$, because one is
prescribing the metric there. On the other hand, one can impose the
Lorentzian condition, that the real part of the second fundamental
form is zero, everywhere else on $\Sigma$. This allows one to find a
saddlepoint solution, bounded by a surface $\Sigma$ with a handle of
area $A$, for any area up to the maximum, $4\pi/\Lambda$. Therefore
the nucleation of Schwarzschild-de~Sitter black hole pairs of any size
can be analysed in the instanton formalism. We choose the cosmological
horizon to be regular in the Euclidean sector, which will lead to a
conical singularity on the black hole horizon. This is allowed as long
as the surface of measurement, $\Sigma$, contains the conical
singularity (since this means that the metric is not varied there).

\begin{figure}[htb]
  \hspace*{\fill}
    \vbox{ \epsfxsize=.8\textwidth \epsfbox{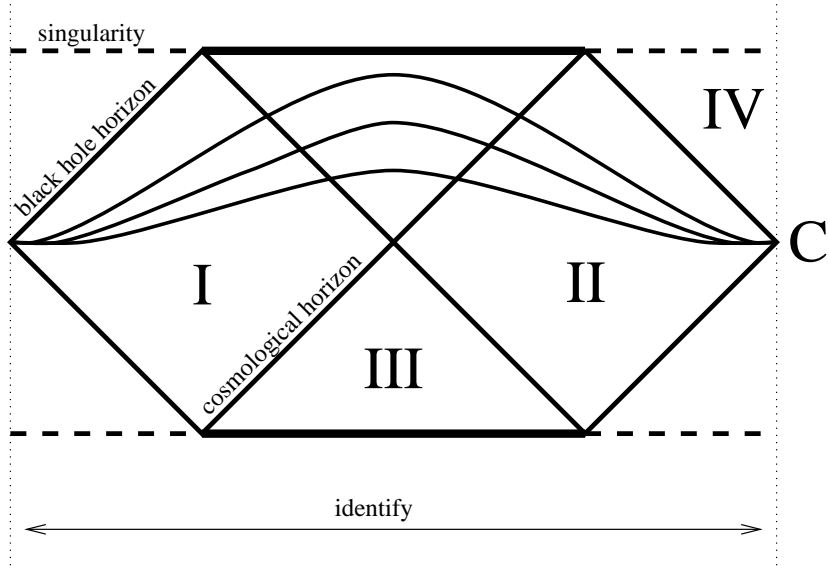}}
  \hspace*{\fill}
\caption[slicing]{\small\sl Carter-Penrose diagram of the
  Schwarzschild-de~Sitter spacetime. The point $C$ is the location of
  the conical singularity in the Euclidean sector. The curved lines
  indicate a family of spacelike slices which all pass through the
  conical singularity. This is necessary since one must specify the
  metric there in order to ensure that the Euclidean solution is a
  saddlepoint. Regions I and II lie between the black hole and the
  cosmological horizon. Region III corresponds to an asymptotic
  de~Sitter region, and region IV to the black hole interior.}
\label{fig-slicing}
\end{figure}

The cross section $\sigma$ corresponds to the black hole horizon; it
will be the smallest $S^2$ in the spacelike surface $\Sigma$. (For
assume it is not. Then the $\sigma$ will not correspond to the conical
singularity, whose metric will then not be fixed on the boundary. But
such configurations will not dominate in the path integral and can be
neglected.)  One can now choose some slicing of
Schwarzschild-de~Sitter which must have the property that the proper
time between points on different slices goes to zero at least
quadratically as a function of proper distance from the black hole
horizon. This type of slicing is shown schematically in a
Carter-Penrose diagram in Fig.~\ref{fig-slicing}.
It ensures that all Lorentzian spacelike slices will be regular on the
black hole horizon. We shall not give any such slicing explicitly.
Once a particular slicing is chosen, there will again be only one
degree of freedom in the second fundamental form, say $K = \int d^3x
h^{1/2} K_{ij} h^{ij}$.

Thus, in the Schwarzschild-de~Sitter case, the wave function has two
arguments, $A$ and $K$. The first determines the size of the black
hole, while the second selects a spacelike slice in the saddlepoint
metric. The de~Sitter and Nariai cases are included for $A=0$ and
$A=4\pi/\Lambda$, respectively.

The Euclidean part of the saddlepoint metric has a boundary with zero
second fundamental form everywhere except on $\sigma$, where it is a
delta function. This boundary will split the full Euclidean solution
in half in the same way as in the de~Sitter and Nariai solutions. This
half of the Euclidean geometry will give the real part of the action.
Choosing $K$ to be purely imaginary leads to a Lorentzian universe,
which once again can be obtained by analytically continuing the
Euclidean solution. Like for the de~Sitter and Nariai solutions, the
Lorentzian section will contribute only to the imaginary part of the
action. Therefore the real part of the action will be independent of
$K$ for imaginary $K$:
\begin{equation}
I_{\rm SdS}(A,K) = I_{\rm SdS}^{\rm Re} (A)
               + i I_{\rm SdS}^{\rm Im} (A,K).
\end{equation}

To calculate the probability measure, and thus the nucleation rate for
a Schwarzschild-de~Sitter black hole pair, we need only calculate the
real part of the action, since
\begin{equation}
\Psi_{\rm SdS}^* \Psi_{\rm SdS} = 
\exp \left[ - 2\, {\rm Re}(I_{\rm SdS}) \right].
\end{equation}
But $2\, {\rm Re}(I_{\rm SdS}) = 2 I_{\rm SdS}^{\rm Re} (A)$, which is
twice the action of the Schwarzschild-de~Sitter instanton, which in
turn is equal to the action of the full Euclidean
Schwarzschild-de~Sitter solution, $I^{\rm full}_{\rm SdS}$.

Using Eq.~(\ref{eq-action}) and $R=4\Lambda$, one can show that
\begin{equation}
I^{\rm full}_{\rm SdS} = - \frac{\Lambda \cal V}{8\pi}
   - \frac{A \delta}{8\pi},
\label{eq-act-vol+cs}
\end{equation}
where $\cal V$ is the four-volume of the Euclidean solution.  The
extra term gives the contribution from a conical deficit angle
$\delta$ at a two-surface of area $A$~\cite{GinPer83}.

In order to facilitate the calculation of this action, it is useful to
parametrize the Schwarzschild-de~Sitter solutions by the radii $b$ and
$c$ of the black hole and the cosmological horizon. The parameters
$\Lambda$ and $\mu$ can be expressed in terms of the new parameters
$b$ and $c$:
\begin{eqnarray}
\Lambda & = & \frac{3}{b^2 + c^2 + bc}
\label{eq-lambdabc} \\
\mu     & = & \frac{bc(b+c)}{2(b^2 + c^2 + bc)}
\label{eq-mubc}
\end{eqnarray}
The Euclidean Schwarzschild-de~Sitter metric is
\begin{equation}
ds^2 = V(r) d\tau^2 + V(r)^{-1} dr^2 + r^2 d\Omega^2,
\end{equation}
where $V(r)$ is given by Eq.~(\ref{eq-v}); in terms of $b$ and $c$ it
takes the form
\begin{equation}
V(r) = \frac{(r-b)(c-r)(r+b+c)}{r (b^2 + c^2 + bc)}.
\label{eq-V-bc}
\end{equation}

To avoid a conical singularity at the cosmological (black hole)
horizon, the Euclidean time $\tau$ must be identified with the period
$\tau^{\rm id}_c$ ($\tau^{\rm id}_b$), where
\begin{equation}
\tau^{\rm id}_{c,\,b} = 2 \pi \left. \sqrt{g_{rr}} \right|_{r=c,\,b}
  \left| \frac{\partial}{\partial r} \sqrt{g_{\tau\tau}}
  \right|_{r=c,\,b}^{-1},
\end{equation}
where $g_{\tau\tau} = 1/g_{rr} = V(r)$. This gives
\begin{equation}
\tau^{\rm id}_{c,\,b} = 4 \pi \left| \frac{\partial V}{\partial r}
\right|_{r=c,\,b}.
\label{eq-tauidbc}
\end{equation}
We choose to get rid of the conical singularity at $r=c$, so the
volume will be
\begin{equation}
{\cal V} = \frac{4\pi}{3} (c^3-b^3) \tau^{\rm id}_{c}.
\label{eq-vol-bc}
\end{equation}
The conical deficit angle at the black hole horizon is by definition
\begin{equation}
\delta = 2 \pi (1 - \frac{\tau_c}{\tau_b});
\end{equation}
the two-sphere area $A$ is obviously $4\pi b^2$.

With $\Lambda$, $\cal V$, $A$, and $\delta$ expressed in terms of $b$
and $c$, Eq.~(\ref{eq-act-vol+cs}) evaluates to:
\begin{equation}
I^{\rm full}_{\rm SdS} = - \pi (b^2+c^2)
\end{equation}
Note that this action is related to the geometric entropy, $S$, and
the total horizon area in the usual
way~\cite{GarGid94,DowGau94b,HawHor95,HawRos95b,ManRos95}:
\begin{equation}
- I = S = \frac{A + A_{\rm c}}{4},
\end{equation}
where $A_{\rm c} = 4 \pi c^2$ is the area of the cosmological horizon.
Thus we obtain for the pair creation rate of arbitrary-size
Schwarzschild-de~Sitter black holes in de~Sitter space:
\begin{equation}
\Gamma_{\rm SdS} = \exp [ - ( I^{\rm full}_{\rm SdS} -
  I^{\rm full}_{\rm dS} ) ] =
\exp (-\pi bc).
\label{eq-pcrsds}
\end{equation}
Using Eqs.~(\ref{eq-lambdabc}) and $A = 4 \pi b^2$, this result can
easily be rewritten in terms of $\Lambda$ and $A$, the argument we
specified in the wavefunction. However, the physical implications are
quite clear from Eq.~(\ref{eq-pcrsds}): a decreasing cosmological
constant corresponds to increasing cosmological horizon size $c$ and
thus, as in the maximal case, to increasing suppression. At fixed
value of the cosmological constant, the suppression increases with the
black hole radius, $b$, which is physically sensible. Considering the
Planck length to be the lower bound on the black hole size ($b \geq
1$), we find that even the smallest black holes are highly suppressed
unless the cosmological constant is also near the Planck value.

Wu has recently proposed~\cite{Cha97} that one should calculate the
saddlepoint approximation to the wave function using ``constrained
instantons'', which include spacetimes with a conical singularity. He
conjectures the conical singularities should be allowed on the
``equator'', i.e.\ the $K_{ij}=0$ surface on which the real Euclidean
geometry is matched to a real Lorentzian one. This is essentially
equivalent to what we have done but the motivation for his calculation
is maybe not so clear. He obtains the same result for the pair
creation probability of sub-maximal black holes.

\sect{Summary and Conclusions}

We have argued that the momentum representation of the wavefunction of
the universe has several advantages over the metric representation.
Most importantly, the requirement that we live in a Lorentzian
universe can be implemented straightforwardly in this formulation: one
must take the argument of the wavefunction to be purely imaginary.
Moreover, unlike the three-metric, the canonical momentum is closely
related to observable quantities like the expansion rate of the
universe, and it distinguishes between expanding and contracting
branches. While the momentum and metric representations are related by
a Laplace transform and thus contain the same information, we conclude
that many of the most relevant physical properties of a spacetime are
manifest only in the momentum representation.

We have clarified how, and under which conditions, Euclidean solutions
with a conical singularity may be used as saddlepoints. We showed that
this is possible in the case of sub-maximal Schwarzschild-de~Sitter
universes if the spacelike boundary, $\Sigma$, is chosen to contain
the conical singularity and the metric is specified there. On the rest
of $\Sigma$, a purely imaginary second fundamental form is specified
to ensure that the observed universe is Lorentzian. This enabled us to
describe the quantum nucleation of such spacetimes and calculate their
creation rate on a de~Sitter background.

\bibliographystyle{board}
\bibliography{all}

\begin{thebibliography}{10}

\bibitem{HarHaw83}
J.~B. Hartle and S.~W. Hawking: {\em Wave function of the {U}niverse\/}. Phys.
  Rev. D {\bf 28}, 2960 (1983).

\bibitem{Vil86}
A.~Vilenkin: {\em Boundary conditions in quantum cosmology\/}. Phys. Rev. D
  {\bf 33}, 3560 (1986).

\bibitem{Lin84b}
A.~Linde: {\em Quantum creation of an inflationary universe\/}. Sov. Phys. JETP
  {\bf 60}, 211 (1984).

\bibitem{GinPer83}
P.~Ginsparg and M.~J. Perry: {\em Semiclassical perdurance of {de S}itter
  space\/}. Nucl. Phys. {\bf B222}, 245 (1983).

\bibitem{BouHaw95}
R.~Bousso and S.~W. Hawking: {\em The probability for primordial black
  holes\/}. Phys. Rev. D {\bf 52}, 5659 (1995), gr-qc/9506047.

\bibitem{BouHaw96}
R.~Bousso and S.~W. Hawking: {\em Pair creation of black holes during
  inflation\/}. Phys. Rev. D {\bf 54}, 6312 (1996), gr-qc/9606052.

\bibitem{GarGid94}
D.~Garfinkle, S.~B. Giddings and A.~Strominger: {\em Entropy in black hole pair
  production\/}. Phys. Rev. D {\bf 49}, 958 (1994), gr-qc/9306023.

\bibitem{DowGau94b}
F.~Dowker, J.~P. Gauntlett, S.~B. Giddings and G.~T. Horowitz: {\em On pair
  creation of extremal black holes and {K}aluza-{K}lein monopoles\/}. Phys.
  Rev. D {\bf 50}, 2662 (1994), hep-th/9312172.

\bibitem{HawHor95}
S.~W. Hawking, G.~T. Horowitz and S.~F. Ross: {\em Entropy, area, and black
  hole pairs\/}. Phys. Rev. D {\bf 51}, 4302 (1995), gr-qc/9409013.

\bibitem{HawRos95b}
S.~W. Hawking and S.~F. Ross: {\em Duality between electric and magnetic black
  holes\/}. Phys. Rev. D {\bf 52}, 5865 (1995), hep-th/9504019.

\bibitem{ManRos95}
R.~B. Mann and S.~F. Ross: {\em Cosmological production of charged black hole
  pairs\/}. Phys. Rev. D {\bf 52}, 2254 (1995), gr-qc/9504015.

\bibitem{Cha97}
W.-Z. Chao: {\em Quantum creation of a black hole\/}. Int. J. Mod. Phys. {\bf
  D6}, 199 (1997), gr-qc/9801020.

\end{thebibliography}

\end{document}